\documentclass[fleqn,a4paper,12pt]{article}
\usepackage{amssymb}
\usepackage{amsmath}
\usepackage[dvips]{graphicx}
\usepackage{epsfig}

\newcommand{\bab}{\end{gather}}
\newcommand{\ri}{{\mathrm i}}
\newcommand{\p}{\partial}
\newcommand{\bea}{\begin{array}}
\newcommand{\eea}{\end{array}}
\newcommand{\beg}{\begin{gather}}

\catcode`@=11 \long
\def\@caption#1[#2]#3{\par\addcontentsline{\csname
ext@#1\endcsname}{#1} {\protect\numberline{\csname
the#1\endcsname}{\ignorespaces #2}} \begingroup \small
\@parboxrestore \@makecaption{\csname fnum@#1\endcsname}
{\ignorespaces #3}\par \endgroup} \catcode`@=12

\newcommand{\la}{\label}

\catcode`@=11 \long
\def\@caption#1[#2]#3{\par\addcontentsline{\csname
ext@#1\endcsname}{#1} {\protect\numberline{\csname
the#1\endcsname}{\ignorespaces #2}} \begingroup \small
\@parboxrestore \@makecaption{\csname fnum@#1\endcsname}
{\ignorespaces #3}\par \endgroup} \catcode`@=12

\begin{document}

\allowdisplaybreaks
 \begin{titlepage} \vskip 2cm

 \begin{center} {\Large\bf Superintegrable quantum mechanical systems with position dependent masses invariant with respect to two parametric Lie groups}

\vskip 3cm {\bf {A. G. Nikitin }\footnote{E-mail:
{\tt nikitin@imath.kiev.ua} }
\vskip 5pt {\sl Institute of Mathematics, National Academy of
Sciences of Ukraine,\\ 3 Tereshchenkivs'ka Street, Kyiv-4, Ukraine,
01024, and\\Universit\'a del Piemonte Orientale,\\
Dipartimento di Scienze e Innovazione Tecnologica,\\
viale T. Michel 11, 15121 Alessandria, Italy}}\end{center}
\vskip .5cm \rm\begin{abstract} { Quantum mechanical
systems with position dependent masses (PDM) admitting two parametric Lie symmetry groups   are classified. Namely,
all PDM  systems are specified which, in addition to their invariance w.r.t. a two
parametric Lie group, admit at least one second order integral of motion. The presented classification is partially extended to the more generic systems which do not accept any Lie group. }
\end{abstract}
\end{titlepage}
\section{Introduction\label{int}}

Symmetry is a corner stone of the majority of modern physical models. It plays a leading role in many research fields, especially in quantum mechanics. The fundamentals of quantum mechanics by definition include symmetries of its basic motion equations which have been discovered in famous papers \cite{Hag}, \cite{Nied}, \cite{And} and \cite{Boy}. We will not discuss these symmetries which are presented in our previous work \cite{AG} but restrict ourselves to note that there are new contemporary results   in this very old field. And it is the case even for classical Lie symmetries \cite{N32, N42, N52}.

The subject of the present paper are symmetries more generic then the Lie ones, namely, the higher order integrals of motion. The systematic study of them had been started  with seminal papers \cite{wint1} were the competed classification of the second order integrals of motion for the 2d Schr\"odinger equation was proposed.  These results induced a great many of generalizations, starting with 3d models \cite{ev1}, \cite{ev2} and continuing with models including matrix potentials \cite{w7, w8,N5} and the higher  (and even arbitrary) order integrals of motion \cite{AGN1, wintL, Tur, N7}. Moreover, such generalizations include the systems more generic than the standard Schr\"odinger equations, namely, Schr\"odinger equations with position dependent mass. The
latter equations are requested in many branches of modern theoretical physics, whose list can be found,
e.g., in   \cite{Roz, Rosas, NZ}. Their symmetries are studied much less than those ones for the standard Schr\"odinder equation. However, the classification of Lie symmetries of the PDM Schr\"odinger equations with scalar potentials have been obtained already \cite{NZ} -\cite{AGN}.

The modern trends are to study the 2d superintegrable systems admitting integrals of motion of
the third and even arbitrary orders \cite{wintL, Tur}.  However, there is only a particular progress in this direction which is restricted to the systems with constant masses and very specific kind of potentials. Nevertheless for the case of the third order integrals of motion such systems with arbitrary potentials have been classified explicitly \cite{Gra1, Gra2} while for the case of the arbitrary order in fact we know only the determining equations which are hardly be solved \cite{Post}. See also \cite{AGN1} where the determining equations for
such symmetries for 3d systems were deduced, and \cite{N7} where symmetry operators of arbitrary order for the
free Schr\"odinger equation had been enumerated.

Second order integrals of motion for 2d  PDM Schr\"odinder equations are perfectly classified  \cite{Kal1, Kal2, Kal3, Kal4}. In particular, it is known that there are 58 inequivalent systems admitting such integrals . The majority of them admits also  at least one continuous  Lie symmetry. Moreover, two dimensional
second-order (maximally) superintegrable systems for Euclidean 2-space even algebraic geometrically \cite{Kress}.

In contrary,  the situation with the 3d systems is not so transparent. The problem of classification of the second order integrals of motion appears to be very complicated. At the best of my knowledge the completed classification results were presented only for the maximally superintegrable (i.e., admitting the maximal possible number of integrals of motion) systems \cite{Kal5, Kal31}, and (or) for the system whose integrals of motion are supposed to satisfy some special condition like  the functionally linearly
dependence \cite{bern}. More exactly, the nondegenerate systems, i.e, those ones which have have 5  linearly independent, contained in 6 linearly independent (but functionally dependent)
2nd order integrals of motion  are known \cite{Cap}, see also \cite{Vol} for the contemporary trends in this field. In addition, a certain progress can be recognized in the classification  of the so called semidegenerate systems which admit only five linearly independent systems and whose potentials are linear combinations of three functionally independent terms \cite{Esco}.   Surely, just such systems are both nice and important since they can be exactly solved and admit solutions in multi separated coordinates \cite{Ra, Ca, Bala2, Rag1}. However, we cannot ignore the PDM systems which admit second order integrals of motion but are not necessary maximally superintegrable and do not belong to quantum analogues of the nondegenerate or semidegenerate ones. And just such systems are studied in the present paper.

Bearing in mind the complexity of the total classification of superintegrable systems with position dependent mass it is reasonable to separate this generic problems to well defined subproblems which can have their own  fundamental and application values. The first subproblem which consists in the  classification of  such systems  admitting  the first order integrals of motion was solved in \cite{NZ}.

An important aspect of the results presented in \cite{NZ} is the complete description of possible Lie symmetry groups which can be admitted by the stationary PDM Schr\"odinger equation. And this property, i.e., the Lie symmetry, can be effectively used to separate the the problem of the classification of the PDM systems admitting  second order integrals of motion for PDM systems to a well defined subproblems corresponding to the fixed symmetries.

As it was shown in \cite{NZ} the PDM Schr\"odinger equation can admit six, four, three, two or one parametric Lie symmetry groups. Surely, there are also such equations which have no Lie symmetry.  In other words, there are six well defined classes of such equations corresponding to these types of symmetries. And it is a natural idea to search for second order integrals of motion consequently for all these classes.

Equations admitting the four or six  parametric Lie groups should have the fixed potential and mass terms, thus the classification of their second order symmetries is a routine problem.

The first informative step was to classify such systems which, in addition to the second order integrals of motion, are invariant with respect to  three parametric Lie groups \cite{AG}. After that we plan to generalize this result to the case when the a priori requested invariance groups are two or at least one parametric. And the final step presupposes the classification of  systems which admit second order integrals of motion but do not have any Lie symmetry.

In the present paper we classify the PDM systems  which admit second order integrals of motion and are invariant with respect to two-parametric Lie groups. Such classification extends  the results of our previous paper \cite{AG} to a much more generic  class of PDM systems.

In spite of the fact that the main stream  in studying of superintegrable systems with PDM is to start with the classical Hamiltonian systems an then quantize them if necessary, we deal directly with quantum mechanical systems. This way is more difficult but guaranties obtaining all integrals of motion including those ones which can disappear in the classical limit \cite{Hit}.

\section{PDM Schr\"odinger equations admitting two parametric Lie groups}

We will  search for superintegrable  stationary  Schr\"odinger equations with position dependent mass of
the following generic form:
\begin{gather}\la{se}
   \hat H \psi=E \psi,
\end{gather}
where
\begin{gather}\la{A1}  H=\frac14(m^\alpha p_a m^{\beta}p_am^\gamma+
m^\gamma p_a m^{\beta}p_am^\alpha)+ \hat V(x)\end{gather}
where $p_a=-i\p_a$, $m=m({\bf x})$ is a function of spatial variables ${\bf x}=(x_1,x_2,x_3)$, associated with the position dependent mass, $\alpha, \beta$  and $\gamma$ are the so called
ambiguity parameters satisfying the condition $\alpha+\beta+\gamma=-1$, and and summation from 1 to 3 is
imposed over the repeating index $a$.

There are various physical speculations how to fix the ambiguity parameters in the particular models based on equation (\ref{A1}) . However, the systems with different values of these parameters are mathematically equivalent up to redefinition of potentials $V({\bf x})$. To simplify the following calculations we will fix them in the following manner: $\beta=0, \gamma=\alpha=-\frac12$ and denote $f=f({\bf  x})=\frac1{2m({\bf x})}$. As a result hamiltonian (\ref{A1}) is reduced to the following form
 \begin{gather}\la{H} H=f^\frac12p_ap_af^\frac12+ V({\bf x}).\end{gather}

In addition we consider the version $\alpha=\gamma=0, \beta=-1$ which corresponds to the following form of hamiltonian (\ref{A1}):
\begin{gather}\la{Ha} H=p_afp_a+ \hat V.\end{gather}

Operators (\ref{Ha}) and (\ref{H}) are equal one to another provided potentials $V=V({\bf x})$ and $\hat V=\hat V({\bf x})$ satisfy the following condition:
\begin{gather}\la{Hb}V=\hat V+V^k\end{gather}
where
\begin{gather}\la{Hc}V^{(k)}=\frac1{4f}((\p_1 f)^2+(\p_2 f)^2+
(\p_3 f)^2)
-\frac12\Delta f\end{gather}
and $\Delta$ is the Laplace operator.

Formula (\ref{Hc}) represents an example of {\it kinematical} potentials which can be reduced to zero by the rearranging  the   ambiguity parameters.

In paper \cite{NZ} all equations (\ref{se}) with Hamiltonians (\ref{Ha})  admitting at least one parametric Lie symmetry group were classified. It was shown that there  are six inequivalent symmetries which can
 be accepted by the PDM Schr\"odinger equations. They include  rotation around the third coordinate axis, shift along this axis
 and dilatation groups. In addition, we can fix three combined symmetries which are superpositions
 of the mentioned ones.

In paper \cite{NZ} all equations (\ref{se}) admitting at least one one-parametric  Lie group were classified.  The list of such equations includes six representatives which accept two-parametric invariance groups. The corresponding inverse masses $f$ and potentials $V$ are presented in the following formulae:
\begin{gather}\la{f_V1}f=F(\tilde r), \quad V=V(\tilde r),\\\la{f_V2}f=\tilde r^2F(\theta), \quad V=G(\theta),\\\la{f_V3}f=\tilde r^2F\left(\frac{r^2+1}{\tilde r}\right),\quad V=G\left(\frac{r^2+1}{\tilde r}\right)\\\la{f_V41}f=\tilde r^2F(a\ln(\tilde r)+\varphi), \quad V=G(a\ln(\tilde r)+\varphi),\\\la{f_V5}f=\tilde r^2F(\varphi), \quad V=G(\varphi),\\\la{f_V6}f=F(x_3), \quad V=G(x_3)\end{gather}
where $F(.)$ and $V(.)$ are arbitrary functions whose arguments are fixed in the brackets,
\begin{gather*} r=(x^2_1+x_2^2+x_3^2)^\frac12,\quad  \tilde r=(x^2_1+x_2^2)^\frac12,\quad \varphi=\arctan\left(\frac{x_2}{x_1}\right), \ \ \theta=\arctan\left(\frac{\tilde r}{x_3}\right).\end{gather*}

Let us stress that functions (\ref{f_V1})-(\ref{f_V6}) are still functions of Cartesian  variables $x_1, x_2$ and $x_3$. However, their dependence on these variables is not arbitrary but rather specific, and it is a consequence of their symmetries.

Equations (\ref{se}), (\ref{H}) whose arbitrary elements are given by formulae (\ref{f_V1}), (\ref{f_V2}), (\ref{f_V3}), (\ref{f_V41}), (\ref{f_V5}) and (\ref{f_V6}) admit the following first order integrals of motion \cite{NZ}:
\begin{gather}\la{IM1}L_3=x_1p_2-x_2p_1, \quad P_3=p_3, \\\la{IM2} L_3, \quad D=x_ap_a-\frac{3\ri}2,\\\la{IM3}P_3-K_3=p_3-r^2p_3+2x_3D, \quad L_3,\\ \la{IM4}P_3, \quad D+\nu L_3,\\\la{IM5}P_3, \quad D\end{gather} and
\begin{gather}\la{IM6}P_1=p_1, \quad  P_2=p_2\end{gather}
correspondingly. These integrals of motion are infinitesimal operators of the inequivalent two parametric Lie groups admitted by the related equations.

Thus the subject of our discussion is a special subclass of PDM Schr\"odinder equations, namely, equations whose arbitrary elements are enumerated in formulae (\ref{IM1})-(\ref{IM6}). They include all inequivalent PDM Schr\"odinder equations admitting two parametric Lie groups. Our task is to specify such of them which, in addition, admit second order integrals of motion.

\section{Determining equations}

Let us search for equation (\ref{se}) which admit  second order integrals of motion, i.e., the second order
differential operators commuting with $H$. Such integrals of motion can be represented in the
following  form:
\begin{equation}\label{Q}
    Q=\p_a\mu^{ab}\p_b+\eta
\end{equation}
where $\mu^{ab}=\mu^{ba}$ and $\eta$ are unknown functions  of $\bf
x$
and summation from 1 to 3 is imposed over all repeating indices.

Operators (\ref{Q}) are formally hermitian. In addition, just representation (\ref{Q}) leads to the most compact and simple systems of determining equations for unknown parameters   $\mu^{ab}$ and $\eta$.

By definition, operators  $Q$ should commute with $H$:
\begin{equation}\label{HQ}[ H,Q]\equiv  H Q-Q H=0.\end{equation}

Our task is to find all inequivalent PDM Hamiltonians  with specific arbitrary elements $f$ and $V$ whose generic form is fixed in (\ref{f_V1}) and (\ref{f_V2}) which admit at least one integral of motion (\ref{Q}) satisfying (\ref{HQ}). As it was noted in \cite{154} it is reasonable to use  representation (\ref{A1}) for the hamiltonian since just this representation leads to the most simple form of the determining equations for symmetries. And this is why we will use it in the following calculations.

Evaluating the commutator in (\ref{HQ})
and equating to zero the coefficients for the linearly independent differential operators $\p_a\p_b\p_c$ and $\p_a$ we come to the following determining equations for arbitrary elements $f, V$ of the Hamiltonian and functions $\mu^{ab}$ , $\eta$ defining integrals of motion (\ref{Q}):
\begin{gather}\la{m0}5\left(\mu^{ab}_c+\mu^{ac}_b+ \mu^{bc}_a\right)=
\delta^{ab}\left(\mu^{nn}_c+2\mu^{cn}_n\right)+
\delta^{bc}\left(\mu^{nn}_a+2\mu^{an}_n\right)+\delta^{ac}
\left(\mu^{nn}_b+2\mu^{bn}_n\right),\\
\la{m1}
 \left(\mu^{nn}_a+2\mu^{na}_n\right)f-
5\mu^{an}f_n=0,\\\la{m2}\mu^{ab}V_b-f\eta_a+F^a=0\end{gather}
  where $\delta^{bc}$ is the Kronecker delta, $f_n=\frac{\p f}{\p x_n}, \ \mu^{an}_n=\frac{\p \mu^{an}}{\p x_n }$, etc., and summation is imposed over the repeating indices $n$ over the values $n=1,2,3$. Moreover,
  \begin{gather}\la{e20b}F^a=(f\mu^{ak}_{km}-\mu^{mn}f_{na})_m\end{gather}
  for the  Hamiltonian of form (\ref{Ha}) and
\begin{gather}\la{e20c}F^a=(f\mu^{ak}_{km}-\mu^{mn}f_{na})_m+\mu^{ab}V^{(k)}_b\end{gather}
for the  Hamiltonian of form (\ref{Hb}).

The term (\ref{e20b}) in general is not trivial. However, it generates a kinematical part of potential. On the other hand the term  (\ref{e20b}) is reduced to zero provided equations (\ref{m0}) and (\ref{m1}) are satisfied. And this is why it is reasonable to use just representation (\ref{Hb}) for the Hamiltinian, since in this case equation (\ref{m2}) is reduced to the maximally simple form, i.e.,
\begin{gather}\la{m2a}\mu^{ab}V_b-f\eta_a=0.\end{gather}

 Thus to classify Hamiltonians (\ref{H}) admitting second order
integrals of motion (\ref{Q}) we are supposed  to find
inequivalent solutions of very complicated system
(\ref{m0})--(\ref{m2}). Its complication is justified in the following speculations.

The autonomous subsystem (\ref{m0}) defines the conformal Killing tensor.
Its general solution is a linear combination of the following tensors
(see, e.g., \cite{Kil})
\begin{gather}\la{K0}
\mu^{ab}_0=\delta^{ab}g({\bf x}),\\
\mu^{ab}_1=\lambda_1^{ab},\\
\la{K1}
\begin{split}&
\mu^{ab}_2=\lambda_2^a x^b+\lambda_2^b x^a-2\delta^{ab}\lambda_3^c
x^c,
\\&
\mu^{ab}_3=(\varepsilon^{acd}\lambda_3^{cb}+ \varepsilon^{bcd}
\lambda_3^{ca})x^d,\end{split}\\
\la{K2}\begin{split}&
\mu^{ab}_4=(x^a\varepsilon^{bcd}+x^b\varepsilon^{acd}) x^c\lambda^d_4,
\\&\mu^{ab}_5=\delta^{ab}r^2+
k (x^ax^b-\delta^{ab}r^2),\\&
\mu^{ab}_6=\lambda_6^{ab}r^2-(x^a\lambda_6^{bc}+x^b\lambda_5^{ac})x^c-
\delta^{ab}\lambda_6^{cd}x^cx^d,\end{split}\\
\la{K3}\begin{split}
& \mu^{ab}_7=(x^a\lambda_7^b+x^b\lambda_7^a)r^2-4x^ax^b\lambda_7^c x^c+
\delta^{ab}
 \lambda_7^c x^cr^2,
\\&\mu^{ab}_8= 2(x^a\varepsilon^{bcd} +x^b\varepsilon^{acd})
\lambda_8^{dn}x^cx^n- (\varepsilon^{ack}\lambda_8^{bk}+
\varepsilon^{bck}\lambda_8^{ak})x^cr^2\end{split}\\
\begin{split}
&\mu^{ab}_9=\lambda_9^{ab}r^4-2(x^a\lambda_9^{bc}+x^b\lambda_9^{ac})x^cr^2+
(4x^ax^b+\delta^{ab}r^2)\lambda_{9}^{cd}x^cx^d\\&+
\delta^{ab}\lambda_{9}^{cd}x^cx^dr^2\end{split}\la{K4}
\end{gather}
where $r=\sqrt{x_1^2+x_2^2+x_3^2}$, $\lambda_m^{ab}=\lambda_m^{ba}$ and  $\lambda_m^a $  are
arbitrary parameters, satisfying the condition $\lambda_m^{nn}=0$, and  $g({\bf x}) $ is an arbitrary function of
$\bf x$.

Thus our  problem presupposes solving the determining equations
(\ref{m1})
and (\ref{m2}) where $\mu^{ab}$ are linear combinations of ten tensors (\ref{K0})-(\ref{K4}). Notice that these tensors include 35 arbitrary parameters in addition to the   coefficients of this linear combination. Moreover, there are four unknown  function, i.e., $f$, $V$ and $g, \eta$. Thus the generic classification of superintegrable systems with position dependent mass looks rather huge. Fortunately, for the systems admitting two parametric continuous symmetry groups specified by equations (\ref{IM1})-(\ref{IM6}) it is possible find all inequivalent solutions of the related determining equations (\ref{m1}) and (\ref{m2}).

  \section{Equivalence relations }

The key element of any classification problem is a clear definition of equivalence relations. This point is especially important in the case of the classification of differential equations whose form is essentially dependent on the chosen variables which we can change and generate infinite  number of equivalent systems.

    Nondegenerated changes of dependent and independent variables of a partial differential equation are called  equivalence transformations provided they keep his  generic form. In our case this generic form is fixed by equations (\ref{se}), (\ref{H}), and, additionally, by relations (\ref{f_V1})-(\ref{f_V6}) if we suppose the invariance with respect to two parametric Lie groups. The equivalence transformations should keep the mentioned generic forms up to the explicit expressions for the arbitrary elements $f$  and $V$.  The have the structure of a continuous group which however can be extended by  some discrete elements.

In accordance with the results presented  in \cite{NZ}, the maximal continuous equivalence group of equation  (\ref{se}) is C(3), i.e., the group of conformal transformations of the 3d Euclidean space. The basis elements of the corresponding Lie algebra can be chosen  in the following form :
\begin{gather}\label{QQ}\begin{split}&
 P^{a}=p^{a}=-i\frac{\partial}{\partial x_{a}},\quad L^{a}=\varepsilon^{abc}x^bp^c, \\&
D=x_n p^n-\frac{3\ri}2,\quad K^{a}=r^2 p^a -2x^aD,\end{split}
\end{gather}
where $r^2=x_1^2+x_2^2+x_3^2$  and $p_a=-i\frac{\p}{\p x_a}.$ Operators $ P^{a},$ $ L^{a},$ $D$ and $ K^{a}$ generate shifts, rotations, dilatations and pure conformal transformations respectively.
The explicit form of these transformations can be found, e.g.,  in \cite{NZ}).

In addition equation (\ref{se}) is form invariant with respect to the following discrete transformations:
 \begin{gather}\la{IT} x_a\to
\tilde x_a=\frac{x_a}{r^2},\quad \psi({\bf x})\to r^3\psi(\tilde{\bf x}).\end{gather}

Notice that algebra c(3) is isomorphic to the algebra so(1,4) whose basic elements $S_{\mu\nu}$ can be expressed via generators  (\ref{QQ}) in the following manner:
 \begin{gather}\la{so} S_{ab}=\varepsilon_{abc}L_c, \quad S_{4a}=\frac12(K_a-P_a),\quad S_{0a}=\frac12(K_a+P_a), \quad S_{04}=D\end{gather}
 where $a, b=1, 2, 3.$
 The related Lie group is SO(1,4), i.e., the Lorentz group in (1+4)-dimensional space. Moreover, the discrete transformation (\ref{IT}) is a realization of the inversion of the fourth coordinate axis.  It anticommutes with $S_{4a}$ but commutes  with the remaining generators (\ref{so}).

 Thus the equivalence group of equations (\ref{se}) with Hamiltonian (\ref{H}) is the conformal group C(3) extended by  discrete transformation (\ref{IT}). This group is locally isomorphic with SO(1,3) extended by the inversion of the fourth  spatial variable.

However, for the systems whose arbitrary elements are fixed by relations (\ref{f_V1})-(\ref{f_V6}) the equivalence group is reduced since it is not admissible to change the invariance groups of these equations. In other words, the set of generators  (\ref{QQ}) should be reduced to the subsets which either commute  with the basis elements of the symmetry algebra presented by relations (\ref{IM1})-(\ref{IM6}) or such commutators are reduced to linear combinations of such elements. In other words, the equivalence group is reduced to the invariance groups which in case (\ref{f_V1}),  and (\ref{f_V5}) is extended by dilatations and in the cases  (\ref{f_V41}), (\ref{f_V6}) by dilatations and rotations around the third coordinate axis. Notice that
 discrete transformation (\ref{IT}) is admissible for the cases  (\ref{f_V1}),  (\ref{f_V3}) and (\ref{f_V5}) only.

\section{Classification results}

Solving the determining equations for all arbitrary elements fixed in  (\ref{f_V1})-(\ref{f_V6}) and applying the equivalence relations discussed in the above we find all inequivalent PDM systems admitting second order integrals of motion, i.e., make the classification of all superintegrable PDM systems which are invariant with respect to two parametric Lie groups. The results of this classification are presented in this section while the calculation details will be given in the following ones.

Let us start with the systems invariant with respect to dilatations and rotations around the third coordinate axis. The generic form of the related inverse masses and potentials are given by equation  (\ref{f_V2}) while the generators of the a priori assumed invariance group are presented in (\ref{IM2}). The more special forms of the inverse masses and potentials which corresponds to the systems admitting second order integrals of motion are represented in the classification tables presented below.

In the classification tables    $F(.), G(.)$ and  $R(.)$ are arbitrary functions of the arguments specified in the brackets,
$ \mu, \nu, \alpha$ and $\kappa$ are arbitrary real parameters,
$\varphi$ and $\theta$ are Euler angles, $ r^2=x_1^2+x_2^2+x_3^2,\ \tilde r^2=x_1^2+x_2^2 $, $
P_a, K_a, D $. The symbol $\{A,B\}$ denotes the anticommutator of
operators  $A$ and $B$, i.e.,
$\{A,B\}=AB+BA.$ In addition, we use the notation
\begin{gather}\la{NN}({F({\bf x})}\cdot H)=p_aF({\bf x})fp_a+ F({\bf x})V.\end{gather}
where $H$ is Hamiltonian (\ref{H}) with arbitrary elements fixed in the second columns of the tables.

In accordance with Table 1 there are five classes of superintegrable PDM systems invariant with respect to dilatations and rotations around the third coordinate axis. They are defined up to  arbitrary parameters and only one of them presented in Item 5 is  maximally superintegrable.

In Items 6 - 9 of the same table we represent the systems  which admit the symmetry with respect to dilatations and shifts along the third coordinate axis. One of them is defined up to two arbitrary functions, the remaining ones  include arbitrary parameters. The systems represented in Items 7-9 are maximally superintegrable.

\newpage

\begin{center}Table 1. Inverse masses, potentials and second order integrals of motion for systems admitting  algebras  $<D, L_3>$  or $<D, P_3>$\end{center}
\begin{tabular}{c c c c c}

\hline

\vspace{1mm}

 No&$f$&$V$&\text{Integrals of motion}&\text{Lie symmetries}\\
\hline\\

1\vspace{2mm}&$\frac{x_3^2 r^2}{ \mu r^2+\lambda x_3^2} $&$\frac{\alpha x_3^2
 +\nu  r^2}{ \mu r^2+\lambda x_3^2}$&
$\begin{array}{c} \{L_1,L_2\}+\left(\frac{2\mu x_1x_2}{ x_3^2}\cdot H \right)-\frac{2\alpha x_1x_2}{ x_3^2},\\\{P_1,K_2\}+\left(\frac{2\lambda x_1x_2}{r^2}\cdot H\right)-\frac{2\alpha x_1x_2}{r^2}
\end{array}$&$D, \ L_3$\\

2\vspace{2mm}&$\frac{r \tilde r^2}{\mu r+\nu x_3}$&$\frac{\alpha r+\omega x_3}{\mu r+\nu x_3}$&$\begin{array}{c}\{P_3,D\}+\left(\frac{\nu}r\cdot H\right)+\frac\omega{r},\\\{D,(K_3\pm P_3)\\-\left(\frac{\lambda(r^2\mp1)}{r}\cdot H\right)+\frac{\omega(r^2\mp1)}{r}
\end{array}$&$D, \ L_3$\\

3\vspace{2mm}&$\frac{x_3^2 \tilde r^2}{ \mu \tilde r^2+\lambda x_3^2} $&$\frac{\alpha x_3^2
 +\nu  \tilde r^2}{ \mu \tilde r^2+\lambda x_3^2}$&
$\begin{array}{c} P_3^2-\left(\frac{\mu }{ x_3^2}\cdot H \right)+\frac{\alpha }{ x_3^2}\end{array}$&$D, \ L_3$\\

4\vspace{1.5mm}&$(r^2\pm1)^2\mp 4x_3^2$&$\alpha$&$\{(K_3\pm P_3),(K_1\mp P_1)\}+30x_3x_1$&$D, \ L_3$\\

5\vspace{1.5mm}&$(r^2\pm1)^2\mp 4x_3^2$&$\alpha$&$\{(K_3\pm P_3),(K_1\mp P_1)\}+30x_3x_1$&$D, \ L_3$\\

5\vspace{1.5mm}&$r^2$&$\alpha$&$\begin{array}{c}\{K_3,P_1\}+\left(\frac{x_3x_1}{r^2}\cdot H\right)\end{array}$&$D, \ L_3, L_1$\\

6\vspace{2mm}&$ \tilde r^2F(\varphi)$&$ F(\varphi)G(\varphi)$&$\begin{array}{c}
L_3^2- (\frac1{F(\varphi)}\cdot H)+G(\varphi)\end{array}$&$D, \ P_3, \ K_3$\\

7\vspace{2mm}&$ \frac{x_1^2x_2^2}{\mu x_1^2+\nu x_2^2}$&$ \frac{\alpha x_1^2+\kappa x_2^2}{\mu x_1^2+\nu x_2^2}$&$\begin{array}{c}
L_3^2-\left (\tilde r^2(\frac\mu{x_2^2}+\frac\nu{x_1^2})\cdot H\right)+\frac{\alpha \tilde r^2}{x_2^2}+\frac{\kappa\tilde r^2}{x_1^2},\\L_1^2-L_2^2+\frac12\left(r^2(\frac\mu{x_1^2}-\frac\nu{x_2^2})\cdot H\right)\\+\frac12r^2(\frac\kappa{x_2^2}-\frac\alpha{x_1^2})\end{array}$&$D, \ P_3, \ K_3$\\

8\vspace{2mm}&$\frac{x_1^2\tilde r}{\mu \tilde r +\nu  x_2}$&$\frac{\alpha  \tilde r+\kappa   x_2}{\mu \tilde r +\nu  x_2}$&$\begin{array}{c} \{P_2,D\}+\{P_3,L_1\}+2\left(\frac{\nu }{\tilde r}\cdot H\right)-2\frac{\kappa  }{\tilde r}, \\
L_3^2-\left(\frac{\tilde r(\mu \tilde r +\nu  x_2)}{x_1^2 }\cdot H\right)+\frac{\tilde r(\alpha   \tilde r+\kappa   x_2)}{x_1^2}\end{array}$&$D, \ P_3, \ K_3$\\

9\vspace{2mm}&$\begin{array}{c}\tilde r^2+\varepsilon x_1 \tilde r,\\\varepsilon=\pm 1\end{array}$&$\frac {\mu x_1+\nu  \tilde r}{\tilde r-\varepsilon x_1}$&$
\begin{array}{c}  \{P_2,D\}+\{P_3,L_1\}-2\left(\frac1{\tilde r}\cdot H\right)-2\frac{\mu}{\tilde r} ,

\\L_3^2- (\frac{\tilde r}{\tilde r+\varepsilon x_1}\cdot H)+\frac{(\mu x_1+\nu  \tilde r)\tilde r}{x_2^2}
\end{array}$&$D, \ P_3, \ K_3$\\

\hline\hline
\end{tabular}

\vspace{3mm}

The next class  of superintegrable systems which we represent in Table 2 are those ones which admit the symmetry with respect to rotations and a specific combinations of the conformal and shift transformations. The related arbitrary elements and generators of the symmetry group are represented by relations (\ref{f_V3}) and (\ref{IM3}). The classification results for such systems are collected in  Table 2.   All of them  are maximally superintegrable.

 The next table, namely, Table 3,  includes three  subclasses of superintegrable systems. The first of them includes the systems invariant with respect to rotations around the third coordinate axis and shifts along this axis. The related inverse masses and potentials are given by equation (\ref{f_V1}) while the generators   of the admissible symmetry group are presented in (\ref{IM1}). The other systems presented in the table  are fixed by relations (\ref{f_V41}),  (\ref{IM4}) and (\ref{f_V6}),  (\ref{IM6}). The systems represented in Items 8 and 9 are maximally superintegrable.

  It is important to note that the second order symmetries presented in the tables are defined up to equivalence transformations discussed in Section 4. In particular, for all systems admitting rotations, i.e., symmetry operators $L_3$ all vector and tensor integrals of motion are defined up to rotations.

 \newpage

 \begin{center}Table 2. Inverse masses, potentials and second order integrals of motion for systems admitting  algebra   $<L_3, K_3-P_3>$\end{center}

\begin{tabular}{c c c c }

\hline

 No&$f$&$V$&\text{Integrals of motion}\\
\hline\\

1\vspace{1.5mm}&$ \frac{\tilde r^2\sqrt{(r^2-1)^2+4x_3^2}}{c_1 \sqrt{(r^2-1)^2+4x_3^2}+c_2(r^2+1)}$&$\frac{c_3 \sqrt{(r^2-1)^2+4x_3^2}+c_4(r^2+1)}{c_1 \sqrt{(r^2-1)^2+4x_3^2}+c_2(r^2+1)}$&$\begin{array}{l}\{D, (K_3-P_3)\}\\-4\left(\frac{c_2}{\sqrt{(r^2-1)^2+4x_3^2}}\cdot H\right)\\+4\frac{c_4}{\sqrt{(r^2-1)^2+4x_3^2}}\end{array}$\\

2\vspace{1.5mm}&$ \frac{(r^2+1)\tilde r^2}{c_1(r^2+1)+c_2\tilde r^2}$&$\frac{c_3(r^2+1)+c_4\tilde r^2}{c_1(r^2+1)+c_2\tilde r^2}$&$\begin{array}{l}\{D,(K_3+ P_3)\}-\frac{2c_4x_3(r^2-1)}{(r^2+1)^2}\\+\left(\frac{2c_2x_3(r^2-1)}{(r^2+1)^2}\cdot H\right),\\
D^2+\left(\frac{c_2r^2}{2(r^2+1)^2}\cdot H\right)-\frac{c_4r^2}{2(r^2+1)^2},\\
(K_1-P_1)^2+(K_2-P_2)^2\\+\left(\frac{4c_2 \tilde r^4-(r^2+1)^4}{2\tilde r^2(r^2+1)^2}\cdot H\right)-\frac{\tilde r(c_2-4)(\alpha c_2-c_4)}{(r^2+1)^2+c_2\tilde r^2}\\+3(3r^2-5x_3^2)\end{array}$\\

3\vspace{1.5mm}&${(r^2+1)^2}$&$\frac{\alpha ( r^2+1)^2}{  \tilde r^2}$&$\begin{array}{c} (K_1-P_1)^2+(K_2-P_2)^2\\+4\left(\frac{\tilde r^2}{(r^2+1)^2}\cdot H\right)+\frac{\alpha(r^2+1)^2}{\tilde r}\\+3(3r^2-5x_3^2),\\\{K_3,P_3\}+\left(\frac{ 2x_3^2}{(r^2+1)^2}\cdot H\right)
\end{array}$\\

4\vspace{1.5mm}&$ \frac{(r^2+1)^2((r^2+1)^2-4\tilde r^2)}{\nu\tilde r^2+\mu(r^2+1)^2}$&$\frac{\alpha (r^2+1)^2-4\kappa \tilde r^2}{\nu\tilde r^2+\mu(r^2+1)^2}$&$\begin{array}{c} (K_1- P_1)^2- (K_2- P_2)^2\\+4(L_2^2-L_1^2)+15(x_1^2-x_2^2)\\+\left(\frac{(4\mu+\nu)( x_2^2-x_1^2)}{ ( r^2+1)^2-4\tilde r^2}\cdot H\right)+\frac{4(\alpha-\kappa)( x_1^2-x_2^2)}{ ( r^2+1)^2-4\tilde r^2}\end{array}$\\

\hline\hline
\end{tabular}

\vspace{3mm}

  For example,  integral of motion $ \{L_1,L_2\}+\left(\frac{2\mu x_1x_2}{ x_3^2}\cdot H \right)-\frac{2\alpha x_1x_2}{ x_3^2}$ presented in Item 1 of Table 1 is a reduced part of the linear combination of the integrals of motion $a Q_{12} + b \tilde Q_{12}$ where
\begin{gather*}\tilde Q_{12}= L_2^2-L_1^2+\left(\frac{\mu (x_2^2-x_1^2)}{ x_3^2}\cdot H
\right)-\frac{\nu (x_2^2-x_1^2)}{ x_3^2}\end{gather*}
and this reduction is made with using the rotation transformations.

In addition we used the discrete equivalence transformation (\ref{IT}) which acts on generators (\ref{QQ}) in the following manner
\begin{gather} P_a\to K_a,\quad K_a\to P_a,\quad L_a \to L_a,\quad D \to D.\la{inv}\end{gather}

The presented tables specify the inequivalent superintegrable systems admitting two parametric Lie symmetry groups. In some cases these symmetry groups are three parametric but include two parametric subgroups, since the related Lie algebras are solvable. To justify the classification  results we will present calculation details in the following sections.

 \newpage

 \begin{center}Table 3.  Inverse masses, potentials and second order integrals of motion for systems admitting  algebras  $<P_3, L_3>, <P_1, P_2>$ or $<P_3, D+\nu L_3 > $\end{center}
\begin{tabular}{c c c c c}
\hline
\vspace{1.5mm}No&$f$&$V$&\text{Integrals of motion}&\text{Lie symmetries}\\
\hline\\

1\vspace{1.5mm}&$\frac{\tilde r^2}{\nu\tilde r^2+\mu}$&$\frac{\alpha\tilde r^2+\lambda}{\nu\tilde r^2+\mu}$&$\begin{array}{c}\{P_3,K_3\}+2\nu(x_3^2\cdot H)-2\alpha x_3^2\end{array}$&$L_3, \ P_3$\\

2\vspace{1.5mm}&$\frac{\tilde r^2}{\mu\ln(\tilde r^2)+\nu}$&$\frac{\alpha\ln(\tilde r^2)+\lambda}{\mu\ln(\tilde r^2)+\nu}$&$L_3D+\mu(\varphi\cdot H)+\alpha\varphi$&$L_3, \ P_3$\\

3\vspace{1.5mm}&$\frac{\tilde r}{\mu \tilde r+\nu}$&$\frac{\alpha \tilde r+\lambda}{\mu \tilde r+\nu}$&$\begin{array}{c}\{P_1,L_3\}-\left(\frac{\nu x_2}{\tilde r}\cdot H\right)+\frac{\lambda x_2}{\tilde r}\end{array}$&$L_3, \ P_3$\\

4\vspace{1.5mm}&$\frac1{\alpha \tilde r^2+\mu}$&$\frac{\nu\tilde r^2+\kappa}{\alpha \tilde r^2+\mu}$&$\begin{array}{c}P_1P_2-\alpha(x_1x_2\cdot H)+\nu x_1x_2,\\
D^2\\-\left(\frac{\alpha}{2r^2}\cdot H\right)+\frac{\nu}{2r^2}\end{array}$&$L_3, \ P_3$\\

6\vspace{1.5mm}&$\frac{\tilde r^2}{\mu(\nu \ln(\tilde r)+\varphi)+\mu}$&$\frac{\alpha(\nu\ln(\tilde r)+\varphi)+\kappa)}{\mu(\nu \ln(\tilde r)+\varphi)+\mu}$&$\begin{array}{c}L_3^2+\alpha\varphi-(\nu\varphi\cdot H)\end{array}$&$P_3, \  L_3 -\nu D$\\

7.&\vspace{2mm}$\frac1{x_3}$&$\frac{c}{x_3}$&$\begin{array}{c}\{P_3,D\}-\frac12((\tilde r^2+4x_3^2)\cdot H)\\-2\alpha x_3,\\
\{P_1,L_1\}-\{P_2,L_2\}\\-(x_1x_2\cdot H)

\end{array}$&$P_1, \ P_2, \ L_3$\\

8.&\vspace{2mm}$\frac{x_3^2}{\mu x_3^2+\nu}$&$\frac{\alpha x_3^2+\lambda}{\mu x_3^2+\nu}$&$\begin{array}{c}\{P_1,K_1\}+2\mu(x_1^2\cdot H)+2\alpha x_1^2,\\

\{P_2,K_2\}+2\mu(x_2^2\cdot H)+2\alpha x_2^2 \end{array}$&$P_1, \ P_2, \ L_3$\\

\hline\hline
\end{tabular}

\vspace{3mm}

\section{Solution of the determining equations}

The determining equations (\ref{m1}) and (\ref{m2}) are nice and look rather gentle. However as it was mentioned in Section 3 in fact they are very complicated systems of partial differential equations including a lot of  unknowns. Fortunately  for the case when the generic functions  $f$ and $V$ are reduced to the forms presented in equations (\ref{f_V1}) we can find the generic solutions of these equations defined up to the equivalence transformations.

The strategy which will be used in solving the determining equations is rather straitforward. The generic solution of equation (\ref{m0}) is known and we represent it in Section 3. The next step is to solve equation  (\ref{m1}) for the inverse mass $f$. To do it directly for the generic Killing tensor represented in Section 3 is absolutely hopeless. However, we can made a priori simplifications of the system (\ref{m1}) by separating it to decoupled subsystems, and this procedure strongly depends on the Lie symmetries accepted by the described systems. In the following subsections we represent the details of such decoupling for all types of symmetries  considered.

Whenever we will obtain the inequivalent versions of the inverse masses, it would be possible to search for solutions of equation (\ref{m2}) for the potential. This step will rather technical and more simple than the previous ones, since the the necessary decoupling of the Killing tensors would be already known.

\subsection{Extended enveloping algebra of c(3)}

Let us start with the note that integrals of motion (\ref{Q}) where $\mu^{ab}$ are linear combinations of the Killing tensors (\ref{K0}) -  (\ref{K4}) can be represented as bilinear combinations of the basic elements of algebra c(3) (\ref{QQ}) added by the special term with $\mu^{ab}=\delta_{ab}g({\bf x})$
and potential term $\eta$. Indeed any of them in fact has the following form
\begin{gather}\la{QQQ}Q=c^{\mu \nu,\lambda\sigma}\{S_{\mu \nu},S_{\lambda \sigma}\}+p_a g({\bf x})p_a\end{gather}
where $S_{\mu \nu}$ are generators (\ref{so}) and $c^{\mu \nu,\lambda\sigma}$ are numeric parameters.

 The conformal Killing tensors (\ref{K0})- (\ref{K4}) are polynomials in $\bf x$ but include also arbitrary functions. For the zero order polynomials (\ref{K0}) representation (\ref{QQQ})  is reduced to the linear combination of products $P_aP_b$, the first order polynomials (\ref{K1}) correspond  to products $P_a L_b$ and $P_a D$, the second order polynomials (\ref{K2}) generate products $P_a K_b, DD $ and $ L_aL_b$ and so on.

 However, equation   (\ref{QQQ}) includes too many terms since the products $\{S_{\mu \nu},S_{\lambda \sigma}\}$ are not necessary linearly independent thanks to certain identities in the extended enveloping algebra of so(3). To avoid possible misunderstandings we present these identities for the bilinear combinations of the basis elements in the following formula:
 \begin{gather}\la{Id}\begin{split}&\{P_a,D\}+\varepsilon_{abc}\{P_b,L_c\}=2P_cx_aP_c, \\& \{L_a,L_b\}+\{P_a,K_b\}=2Q^{ab},\quad  a\neq b,\\&
\{P_1,K_1\}+\{ P_2,K_2\}+L_3^2=2Q^{33},\\&\{K_\alpha,P_\alpha\}+2L_3^2+2D^2=2P_a(r^2-x_\alpha^2) P_a,\\&P_aL_a=0,\\&\{P_a,K_a\}=-4D^2+2P_ar^2P_a,\\&L_1^2+L_2^2+L_3^2
=P_ar^2P_a-D^2,\\&\{P_a,K_b\}-\{P_b,K_a\}=2\varepsilon_{abc}L_cD,\\&P_1^2+P_2^2=-P_3^2+P_aP_a
\end{split}\end{gather}
where $Q^{ab}=P_cx_ax_bP_c, \ \alpha=1, 2, 3$ and no sum w.r.t. $\alpha$.

We  use relations   (\ref{Id}) to produce maximally  compact presentations for the integrals of motion.
\subsection{PDM systems admitting dilatation}
In accordance with   (\ref{IM1}) - (\ref{IM6}) the half part of the considered systems admit the dilatation as an equivalence transformation. Moreover, two of these systems  possess the symmetry with respect to the dilatation transformations whose generator $D$ is presented in (\ref{IM2}). This property enables essentially simplify the solution of the determining equation in accordance with the following reasons.

For the case of the scale invariant systems  admitting second order integrals of motion
the related Killing tensors cannot include
linear combinations of all polynomials listed in  (\ref{K0})-(\ref{K4})  but are reduced to homogeneous
polynomials. Indeed, under the dilatation transformation $x_a\to \alpha x_a$ operators (\ref{Q}) including zero order Killing tensor (\ref{K0}) obtains the multiplier $-\alpha^2$, for case of the first order Killing tensors  (\ref{K1})  we obtain the multiplier $-\alpha,$ etc. In other words, the determining equations     (\ref{m1})
and (\ref{m2}) are reduced to the five  decoupled subsystems corresponding to the Killing  tensors
which are $n$-order homogeneous polynomials with $n=0, 1, 2, 3, 4$, and arbitrary functions
$g_1, g_2, ...,g_9$ should satisfy the following conditions:
\begin{gather}\la{vp}x_ag({\bf x})_a=ng({\bf x}).\end{gather}
Moreover, since Hamiltonians
(\ref{H}) with arbitrary elements
(\ref{f_V3}) and (\ref{f_V5}) are invariant with respect to the inverse transformation (\ref{IT})
 we can restrict ourselves to the polynomials of order $n < 3$, since symmetries with $n$=3 and
 $n$=4 appears to be equivalent to ones with $n=1$ and $n=0$ correspondingly. In other words, it is sufficient to solve the determining equations for the case when the conformal Killing tensors are given by relations (\ref{K0}), (\ref{K2}) and (\ref{K2}), moreover, to do it separately for all the mentioned tensors.

Let us  search for the inverse mass functions $f$bsatisfying equations (\ref{m1}).
For the systems invariant w.r.t. the dilatation transformations these functions $f$ satisfies one more condition
\begin{gather}\la{NONO}x_af_a=2f\end{gather}
which is obviously correct in view of (\ref{f_V2}) and (\ref{f_V2}). However this identity makes it possible to reduce (\ref{m1}) to the
following    {\it homogeneous} system of linear algebraic equations for derivatives   $f_a$:
\begin{gather}\la{NO}M^{ab}f_b=0\end{gather}
where
\begin{gather*}\begin{split}& M^{ab}=\mu^{ab},\\&  M^{ab}=\mu^{ab}-\lambda^ax^b-\mu^ax_b\end{split}\end{gather*}
and
\begin{gather} \la{deq51}M^{ab}=\mu^{ab} -\lambda^{ac}x_cx_b\end{gather}
for Killing vectors (\ref{K0}), (\ref{K1}) and (\ref{K2}) correspondingly.

Let us note  that  for the Killing tensors (\ref{K0}) and (\ref{K1}) functions  $g({\bf x})$
can be expressed via $f$, namely:
\begin{gather}\la{vp1}g({\bf x})=\frac1{2f}{-x_a M^{ab}f_b}\end{gather}
and
\begin{gather}\la{vp2}g({\bf x})=\frac1{f}{-x_a M^{ab}f_b}{f}\end{gather}
correspondingly, while for the tensors (\ref{K3}) we have:
\begin{gather}\la{vp3}g({\bf x})=fG(\varphi,\theta)\end{gather} where $ G(\varphi,\theta)$ is yet unknown function of Euler angles satisfying the equation
\begin{gather}\la{vp4}G_\varphi=\frac1{f^2}(x_aM^{bc}f_c-x_bM^{ac}f_c)\end{gather}

Equations (\ref{vp1}) - (\ref{vp4}) are algebraic consequences of (\ref{vp}), (\ref{NONO}) and (\ref{NO}) obtained by multiplication on $x_a$ and summing up with respect to the repeating index $a$.

Equation (\ref{NO}) admits nontrivial solution iff the determinant of the matrix whose entries are $M^{ab}$
 is equal to zero. It is necessary to specify the admissible combinations of arbitrary constants which correspond to the trivial determinants and than find solutions of the corresponding equations (\ref{m1}) and (\ref{m2}).

Thus the symmetry with respect to the dilatation transformations makes it possible essentially simplify the classification procedure.

\subsection{ PDM admitting rotations}

As it is fixed in (\ref{IM1}) and (\ref{IM2}) the two classes of the considered system are invariant with restpect to the one parametric rotation group. This property  also helps to simplify the classification procedure. Namely, we can decouple the determining equations and consider separately the integrals of motion which are scalars, vectors and tensors with respect to these rotations. Moreover, for the cases of vector and tensor integrals of motion  it is sufficient to specify  only one out of two components of them.

The scalar integrals of motion have to commute with $L_3$. Applying this restriction to the generic bilinear form  (\ref{QQQ}) we can specify the following scalars:
\begin{itemize}
\item
 Bilinear combinations of $S_{12}, \ S_{43}, \ S_{03}, \ S_{04}$;
 \item Linear combinations of the scalar products  $ S_{n1}S_{m1}+S_{n2}S_{m2}$ with $n, m = 3, 4, 0$;
 \item Skew symmetric products $S_{n1}S_{m2}-S_{n2}S_{m1}$
\end{itemize}

It is easy to construct also the vector and tensor combinations. The vector  components are linear combinations of the products $S_{nm}S_{ka}$ with $n, m, k=3,4,0$ and $a=1,2$. In addition,we can set $n=1, m=2$ . The tensor components  look as $S_{n1} S_{m2}+S_{n2} S_{m1}$ and $S_{n1} S_{m1}-S_{n2} S_{m2}$.

Thus the rotation invariance helps to decouple the determining equations to three subsystems and reduce their   number considering only one component of the vector and tensor equations. The number of the arbitrary parameters can be additionally reduced using  identities  (\ref{Id}), but we still have a lot of them. Fortunately, we deal with the systems admitting two parametric Lie groups, and any of them generate their own reduction.  In particular, the systems with the inverse masses and potentials of generic form (\ref{f_V5}) admit both the rotation and dilatation symmetries, and so we can use both the tools presented in this and previous subsections. As a result we were able to discover the systems presented in Items 1-10 of Table 1.  In addition, the systems fixed in (\ref{f_V3}) admit two rotations, one of them on the plane 1-2 and the other on plane 3 -4. Any of these symmetries  generate its own decoupling, and the related integrals of motion are subdivided to scalar-scalar, scalar-vector, vector-scalar, vector-vector scalar-tensor, vector-tensor, tensor-vector, tensor-scalar, tensor-vector and tensor-tensor ones, so the decoupling is very essential. Using it we find the integrals of motion presented in Table 2.

In the following subsection we discuss  the tools presented by the symmetry with respect to translations along two of  coordinate axis.

\subsection{PDM systems invariant with respect to shifts}

The last symmetry we discus is generated by operators $P_a$.
Since the related arbitrary elements (\ref{f_V6}) do not depend on $x_1$ and $x_2$  it is possible essentially reduce the number of admissible  second order integrals of motion.

Let $Q$ be an integral of motion (\ref{Q}) admitted by equation (\ref{se}) with arbitrary elements (\ref{f_V6}).  By definition $P_a$
with $a=1, 2$  are integrals of motion too, the same is true for the commutators $[P_a,Q]$,
$[P_a,[P_b,Q]$ and $[P_a,[P_b,[P_c,Q]]$.  Thus any second order symmetry induces the symmetry  generated by $\mu_1^{ab}$ (refer
to (\ref{K1}), i.e.,
\begin{gather}\la{a0}Q=P_a(\lambda_{ab}+\delta_{ab}g({\bf x}))P_b+\eta.\end{gather}
Moreover, there are the following qualitatively different versions of coefficients $\lambda^{a b}$ in (\ref{a0}):
\begin{gather}\la{a1}\lambda^{3\mu}\neq0, \ \lambda^{\mu\nu}=\lambda^{33}=0, \  \mu, \nu=1,2,\\ \la{a2}
 \lambda^{3\mu}=0,  \text{\  some of coefficients \ } \lambda^{\mu\nu} \text{\ or \ } \lambda^{33} \text{\ are nontrivial} \end{gather}
 which correspond to the following integrals of motion:
 \begin{gather}\la{a3} Q^{31}=P_3P_1+\eta^{31}, \ \ Q^{32}=P_3P_2+\eta^{32}.\end{gather}

In the case (\ref{a2}) the related integrals of motion  are trivial since the  Hamiltonians considered in this section commute with them by definition.

In the case (\ref{a2}) we have to consider  integrals of motion whose commutators with $P_1$ and $P_2$ are reduced to  $Q^{\mu\nu}$ with $\mu, \nu<3$ and  $Q^{33}$. They are listed in the following formula:
 \begin{gather}\la{a5} Q^{a}=\{P_a,D\}+\eta^a,\ \ \tilde Q^a=\{P_3,L_a\}+\tilde \eta^a.\end{gather}

 Thus the superintegrable PDM systems which are invariant with respect to algebra whose basis elements are presented in (\ref{IM6}) have  to admit the integrals (\ref{a3}) and  (\ref{a5}). Such systems can admit some additional symmetries whose calculation for {\it known} system is a rather simple problem.

\subsection{Selected   calculations}
In the previous subsection we specify the algorithms used for solution of the determining equations and show how the a priori requested symmetries can be used for the optimization of the calculations. These algorithms make it possible essentially reduce the volume of calculations. Nevertheless, the number of inequivalent systems of the determining equations which we have to solve is still rather extended. We will not reproduce all of them but restrict ourselves to some important examples.

The most universal integral of motion looks as:
\begin{gather}\la{L^2}Q_1=L_3^2+P_agP_a+\eta.\end{gather}
Such integrals can be admitted by all the considered systems, though for the systems admitting $L_3$ it is  trivial.

The nonzero entries of the Killing tensor corresponding to (\ref{L^2}) are
\begin{gather}\la{NN1}\mu^{11}=x_2^2, \quad \mu^{22}=x_1^2,\quad  \mu^{12}= \mu^{21}=-x_1x_2\end{gather}
while the related determining equations (\ref{m1}) and (\ref{m2}) are reduced to the following forms:
\begin{gather}\la{s1}\begin{split}&x_2f_\varphi-gf_1+fg_1=0,\\&x_1f_\varphi+gf_2-fg_2=0,\\&gf_3-fg_3=0
\end{split}\end{gather}
where $f_\varphi=\frac{\p f}{\p \varphi}$ and $f_a=\frac{\p f}{\p x_a}$, and
\begin{gather}\la{NN2}\begin{split}&x_2V_\varphi-gV_1+f\eta_1=0,\\&x_1V_\varphi+gV_2-f\eta_2=0,
\\&gf_3-f\eta_3=0
\end{split}\end{gather}
Let us remind that the unknowns $f$ and $g$ are connected by relations  (\ref{vp3}) and (\ref{vp4}), which are reduced to the following form:
\begin{gather}\la{vpp}g({\bf x})=fG(\varphi)\end{gather} where $ G(\varphi)$  a function of the Euler angle. Substituting (\ref{vpp}) into equations (\ref{s1}) and integrating them we obtain: \begin{gather}\la{vp44}f=\frac{\tilde r^2}{\tilde r^2F(x_3,\tilde r)-G(\varphi)}.\end{gather}

The next step is to substitute (\ref{vp44}) into (\ref{NN2}) and to solve the obtained system. As a result we obtain the following expressions for $V$ and $\eta$:
\begin{gather}\la{vp5}V=\frac{N(\varphi)+M(\tilde r,x_3)}{G(\varphi)-\tilde r^2F(\tilde r,x_3)}, \quad \eta= \frac{\tilde r^2F(\tilde r,x_3)(N(\varphi)+M(\tilde r,x_3))}{G(\varphi)-\tilde r^2F(\tilde r,x_3)}\end{gather} which make it possible to reduce the integral of motion (\ref{L^2}) to the following form
\begin{gather}\la{vp6}Q_1=L_3^2+(G(\varphi)\cdot H)+N(\varphi)+M(\tilde r,x_3).\end{gather}

The obtained results are valid   for generic PDM system admitting the integral of motion of the form presented in (\ref{L^2}). For the special  types of such systems admitting two parametric Lie groups they are reduced to the versions represented in Items 6-10 of Table 1 and Items 5, 6 of Table 3. The corresponding potential and functions $\eta$ are easily calculated by integrating equations (\ref{NN2}) with known $f$.

 Notice that if only the dilatation symmetry is requested, function (\ref{vp4}) is reduced to the following form:
 \begin{gather}\la{s3}f=\frac{\tilde r^2}{F(\varphi)+G(\theta)}, \end{gather}
 while the related functions $V$ and $\eta$ are \cite{AG}:
 \begin{gather}\la{s4}V=\frac{R(\theta)+
N(\varphi)}{F(\varphi)+G(\theta)}, \quad \eta= -F(\varphi)V+G(\theta).\end{gather}

The next (and the last) example which we consider is the simplest integral of motion whose conformal Killing tensor is a constant added by a diagonal term:
\begin{gather}\la{P^2}Q=P_3^2+P_agP_a+\eta\end{gather}
The related matrix $\mu^{ab}$ has the only nonzero entry $\mu^{33}=1$ which generate the following equations  (\ref{m1}) and (\ref{m2}) :
\begin{gather}\la{P^2a}g f_a=f g_a, a=1,2,\\\la{P^2b}  (g+1)f_3=f g_3,\\\la{P^2c} gV_a=f \eta_a, \quad (g+1) V_3=f \eta_3\end{gather}

In accordance with (\ref{P^2a}) $g=-f G(x_3)$ and so the generic solution of (\ref{P^2b}) is
\begin{gather}\la{P^2e}f=\frac1{F(x_1,x_2)+G(x_3)}\end{gather}

Substituting the obtained expression for $f$ into (\ref{P^2c}) and integrating the latter system we obtain
\begin{gather}\la{P^2d}V=\frac{M(x_1,x_2)+N(x_3)}{F(x_1,x_2)+G(x_3)},  \quad \eta=-G(x_3)V+N(x_3).\end{gather}
where $c$ is the integration constant. The related integral of motion (\ref{P^2}) takes the following form:
\begin{gather}\la{last!}Q_2=P_3^2-\left(G(x_3)\cdot H\right)+N(x_3)\end{gather}
and is valid for arbitrary PDM system (\ref{se}) with rather generic inverse mass and potential  presented in (\ref{P^2e}), (\ref{P^2d}) since we did not ask for any Lie symmetry. Surely it is valid for some of the particular systems enumerated in (\ref{f_V1})-(\ref{f_V6}), see Items 3, 4, and   10 of Table 1 were the integrals of motion including $P_3^2$ and $P_1^2$ are presented.

We see that the calculations requested for solution of particular sets of the determining equations are not too complicated provided one uses the tools outlined in the previous subsections.

\section{Discussion}

In the present paper we continue the procedure of the complete classification of superintegrable quantum mechanical systems with position dependent masses, started in \cite{AG} where the first order constants of motion were found, and  \cite{154}   where the systems  admitting three parametric Lie groups were classified. Now we are presenting the
inequivalent PDM  systems  which admit second order integrals of motion and  two parametric symmetry groups. The total number of such systems is equal to twenty one. One of them include  arbitrary functions while  the remaining ones are defined up to arbitrary parameters.

The number of the presented systems cannot be reduced if we extend the equivalence relations discussed in Section 4 by  St\"ackel transformations, see, e.g., \cite{St} for exact definitions. We will  not discuss this point in details but mention that the St\"ackel equivalent systems can be identified by the similarity of their integrals of motion which however  can have different terms $(g\cdot H)$. The only simplification which can be obtained applying the St\"ackel transform to the presented PDM system is a possible reduction of the number of arbitrary parameters.

All the presented systems are superintegrable, but only nine   of them possess the maximal superintegrability.  The advantage of our approach is that we were  able to find all systems admitting second order integrals of motion including those ones which are not maximally superintegrable.

 We present all inequivalent integrals of motion admitted by the systems under study. The number of linearly independent integrals of motion is more extended, but all of them can be found using the equivalence relations fixed in Section 4. Moreover, we represent just such integrals of motion  which are necessary to be able to find the remaining ones using the a priori fixed symmetries of the systems under study. In contrary, in paper \cite{AG} all linearly independent integrals of motion are represented explicitly.

 It would be interesting to study the algebraic properties of the found integrals of motion. Like in the case of nondegenerate or semidegenerate classical systems \cite{Esco} they generate polynomial algebras. The analysis of these algebras is one of the challenges created by the present paper.

The next natural steps are to classify such the mentioned systems which admit at least one parametric continuous symmetry group, and the systems which do not have any Lie symmetry. Some elements of such classification can be found in paper \cite{AG} where the systems admitting the dilatation symmetry were studied. However, the classification presented in \cite{AG} was restricted to the integrals of motion which, up to potential terms, belong to the standard (non extended) enveloping algebra of c(3).

 We plane to complete the classification of the superintegrable systems admitting dilatation and to classify the systems admitting the other one parametric groups. Notice that in accordance with the results of paper \cite{NZ}
there exist five  inequivalent Lie groups which can be accepted by the 3d
quantum mechanical systems with PDM. Moreover, the superintegrable systems admitting one of this groups are preliminary classified in \cite{fin}.

The classification of the PDM systems which do not posses any Lie symmetry but admit second order integrals of motion would finish the completed description of the superintegrable PDM Schr\"odinger equations. As we mentioned in Introduction this problem looks to be very complicated. However, in spite of the absence of various points generated by the a priori requested symmetries there are some  advantages just for the symmetry less systems since the related equivalence group is maximally extended.

Notice that in the present paper we demonstrate two examples of the superintegrable PDM systems which do not posses any Lie symmetry, see equations (\ref {vpp}),  (\ref {vp5}), (\ref {vp6}) and (\ref {P^2e}), (\ref {P^2d}), (\ref {last!}) in the above. And there is the challenge to make the complete classification of all such systems which we have accepted.

\vspace{2mm}

{\bf Acknowledgement} I am indebted with Universit\'a del Piemonte Orientale and Dipartimento di Scienze e Innovazione Tecnologica for the extended stay as Visiting Professor.

\end{document}